\documentclass{article}

\pdfoutput=1

\usepackage{arxiv}

\usepackage[utf8]{inputenc} 
\usepackage[T1]{fontenc}    
\usepackage{hyperref}       
\usepackage{url}            
\usepackage{booktabs}       
\usepackage{amsfonts}       
\usepackage{nicefrac}       
\usepackage{microtype}      
\usepackage{graphicx}       
\usepackage{physics}        

\title{Dynamic neutron imaging of argon bubble flow in liquid gallium in external magnetic field}

\author{
  Mihails Birjukovs\\
  Institute of Numerical Modelling\\
  University of Latvia\\
  Riga, Latvia, Jelgavas 3, 1004 \\
  \texttt{mihails.birjukovs@lu.lv} \\
   \And
 Valters Dzelme\\
  Institute of Numerical Modelling\\
  University of Latvia\\
  Riga, Latvia, Jelgavas 3, 1004 \\
  \texttt{valters.dzelme@lu.lv} \\
  \And
 Andris Jakovics\\
  Institute of Numerical Modelling\\
  University of Latvia\\
  Riga, Latvia, Jelgavas 3, 1004 \\
  \texttt{andris.jakovics@lu.lv} \\
  \And
 Knud Thomsen \\
  Research with Neutrons and Muons\\
  Paul Scherrer Institut\\
  Villigen, Switzerland, Forschungsstrasse 111, 5232 \\
  \texttt{knud.thomsen@psi.ch} \\
   \And
 Pavel Trtik \\
  Research with Neutrons and Muons\\
  Paul Scherrer Institut\\
  Villigen, Switzerland, Forschungsstrasse 111, 5232 \\
  \texttt{pavel.trtik@psi.ch} \\
}

\begin{document}

\maketitle

\begin{abstract}

This paper presents detailed results of neutron imaging of argon bubble flows in a rectangular liquid gallium vessel with and without the application of external horizontal magnetic field. The developed image processing algorithm is presented and its capability to extract physical information from images of low signal-to-noise ratio is demonstrated. Bubble parameters, velocity components, trajectories and relevant statistics were computed and analysed. A simpler version of the code was applied to the output of computational fluid dynamics simulations that reproduced the experiment. This work serves to further validate the neutron radiography as a suitable method for monitoring gas bubble flow in liquid metals, as well as to outline procedures that might help others to extract data from neutron radiography images with a low signal-to-noise ratio resulting from high frame rate acquisitions required to resolve rapid bubble motion.

\end{abstract}


\section{Introduction}

Gas bubble flow through liquid metal in presence of static external magnetic field is of great interest, because this type of flow occurs during liquid metal stirring, purification, homogenization and crystallization processes, as well as in liquid metal column reactors \cite{r1,r2}. Flow parameters must be tailored to each process, which requires the means for reliable control. To this end, application of static magnetic field has been proposed. Theoretical considerations and experimental evidence indicate that, depending on field orientation, one could achieve bubble jet stabilization and/or alter bubble velocity \cite{r3,r4}. 

Optimization of industrial processes requires both simulations and suitable experimental methods. Neutron radiography is, in the context of high frame rate flow monitoring, a relatively new and promising technique that can probe optically opaque liquid metals directly \cite{r5} in the sense that, in addition to qualitative observations and image-based velocimetry, it enables the study of bubble shape dynamics and characteristic oscillation frequencies, which are important comparison criteria for verification of computational fluid dynamics (CFD) models. For some higher-Z metals, neutron imaging, being conceptually similar to X-ray imaging, is more appropriate, enabling one to work with thicker (i.e. more representative) samples \cite{r5,r6}. This is especially important for suppressing the influence of bubble-wall interactions. Thus, neutron radiography enables one to probe a broader class of multiphase flow systems \cite{r7,r8}, which motivates further validation of this approach as a reliable technique for verification of simulation results and for multiphase flow analysis. 

To date, several noteworthy studies employing neutron radiography have been conducted, but most were focused on single phase flow analysis using tracer particles \cite{r5,r6}, solidification dynamics \cite{r6}, or performed only qualitative analysis of bubble flow. The latter was rather restricted, mainly due to the challenges stemming from the low quality of obtained images [6]. Notable exceptions include the work by Zboray et al. \cite{r8,r9,r10}, wherein image processing and phase tracking algorithms were implemented and flow assessed by neutron radiography was given a quantitative treatment. However, quantitative assessment of gas bubbles in the gas/liquid metal flow context is scarce. The objective of the present study was to acquire sequences of high temporal resolution neutron radiographies of argon (Ar) bubble flow in liquid gallium (Ga) and, for the first time, to perform image-based velocimetry and shape analysis for different Ar flow rates, with and without applied horizontal magnetic field.

\section{The experiment}
\label{sec:experiment}

\subsection{The MHD system}

\begin{figure}[htbp]
\centering
\includegraphics[width=\textwidth]{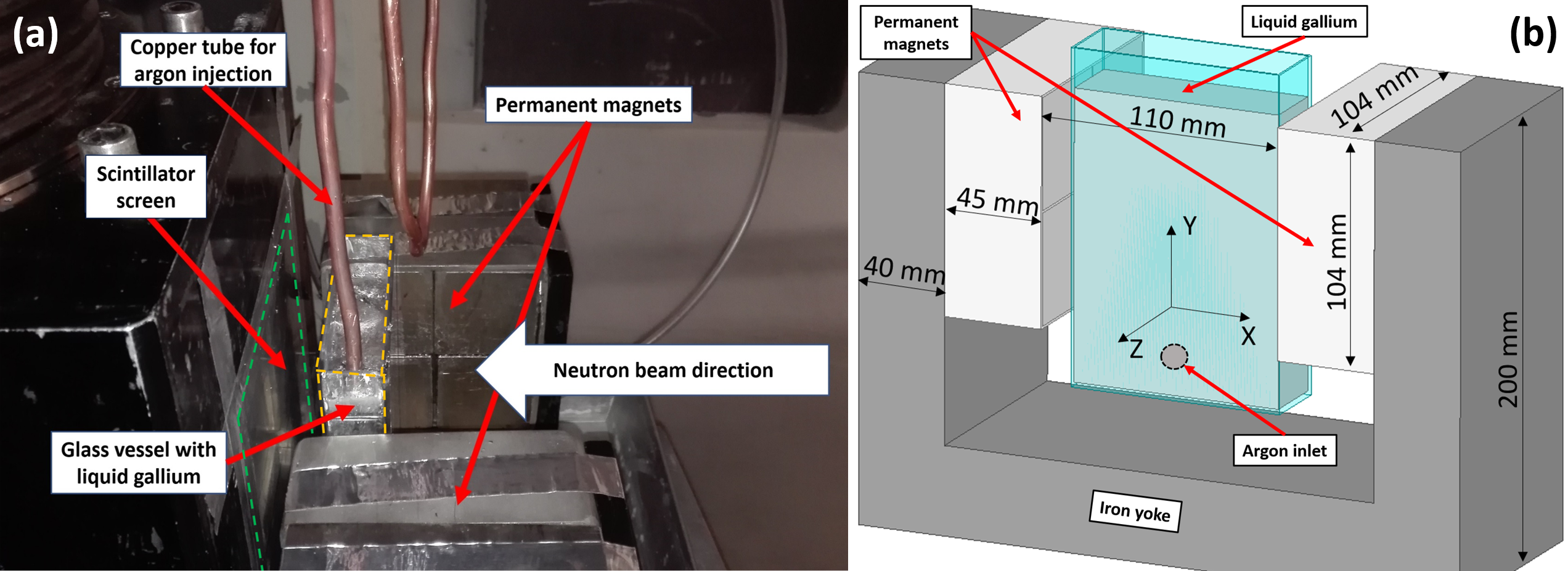}
\caption{(a) Photograph of the experimental setup and (b) its simplified representation with highlighted dimensions.}
\label{fig:setup}
\end{figure}

Experiments were conducted at the thermal neutron imaging beamline \emph{NEUTRA} at the Paul Scherrer Institute PSI \cite{r11}. The setup (Figure \ref{fig:setup}a) consisted of a thin-walled glass container filled with liquid Ga, wherein Ar bubble flow was introduced via a submerged copper tube with a constricted nozzle, which ejected bubbles such that they ascended without wall interactions. The tube was bent so that the outlet was directed horizontally, yielding a useful effect – for a given flow rate, bubble size was constant regardless of nozzle constriction, and was prescribed by Ar/Ga surface tension. The container was connected to a pressurized Ar vessel and flow rate was measured and controlled using a digital mass flow controller (\emph{MKS Instruments 1179B}).

\begin{figure}[htbp]
\centering
\includegraphics[width=\textwidth]{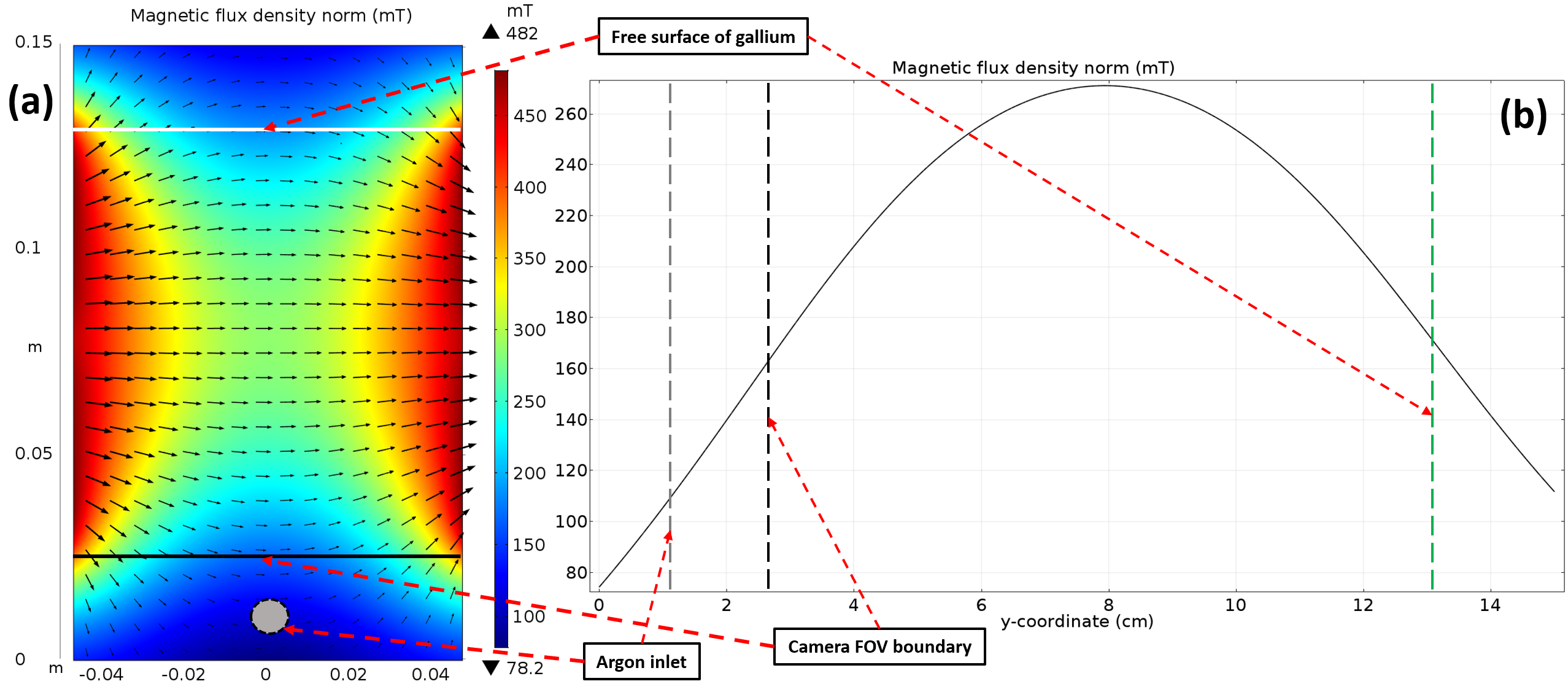}
\caption{Simulated magnetic field (a) within the melt container and (b) along its vertical axis.}
\label{fig:magfield}
\end{figure}

To study the influence of applied horizontal magnetic field, the container was placed between two arrays of neodymium permanent magnets (Figure \ref{fig:setup}b). Magnetic field flux density within the container (Figure \ref{fig:magfield}) ranged from $60~mT$ to $500~mT$ and was roughly $300~mT$ within the bubble flow region, as determined by simulations and measurements. As shown in Figure \ref{fig:magfield}a, the direction of the field was rather homogeneous along container vertical axis and did not exhibit too strong magnitude variations, as seen in Figure \ref{fig:magfield}b.

\subsection{Neutron imaging}

The experimental setup was imaged at the measuring position $2$ of the \emph{NEUTRA} beamline using a medium spatial resolution set-up (MIDI). After passing through the sample, the attenuated neutron beam was detected by $200 ~ \mu m$ thick ${}^6$\emph{LiF/ZnS} scintillator screen. The distance between the centre of the liquid Ga vessel and the detector was $32 ~ mm$. A sCMOS camera (\emph{ORCA Flash 4.0}) was used to collect the scintillator light output. The utilized detector optics produced images with isotropic pixel size of $55.1 ~ \mu m$, with a field of view (FOV) of $112.8$ x $112.8 ~ mm$. As such, the FOV was large enough to capture the entire trajectory of a bubble once it detached from the inlet. 

All images were acquired with a high frame rate – $100$ frames per second (FPS) – to capture the motion of ascending bubbles in detail. Bubble trajectories were recorded for $10$ to $300 ~cm^3/min$ flow rates, with and without applied magnetic field. $30~s$ sequences were acquired for all flow rates, resulting in $3000$ images per measurement. Images were recorded for two types of $95$ x $150 ~mm$ glass containers, $20$- or $30~mm$ liquid Ga thickness.

\section{Image Processing}
Preprocessing was performed in \emph{ImageJ}: pixels with exceedingly high intensities were removed and $4000$ frame average dark current signal was subtracted from raw images, followed by normalization with respect to the $8000$ image average of the unobstructed beam signal to compensate for neutron beam non-uniformity. Afterwards, the images were imported into \emph{Wolfram Mathematica 12} for post-processing.

\begin{figure}[htbp]
\centering
\includegraphics[width=\textwidth]{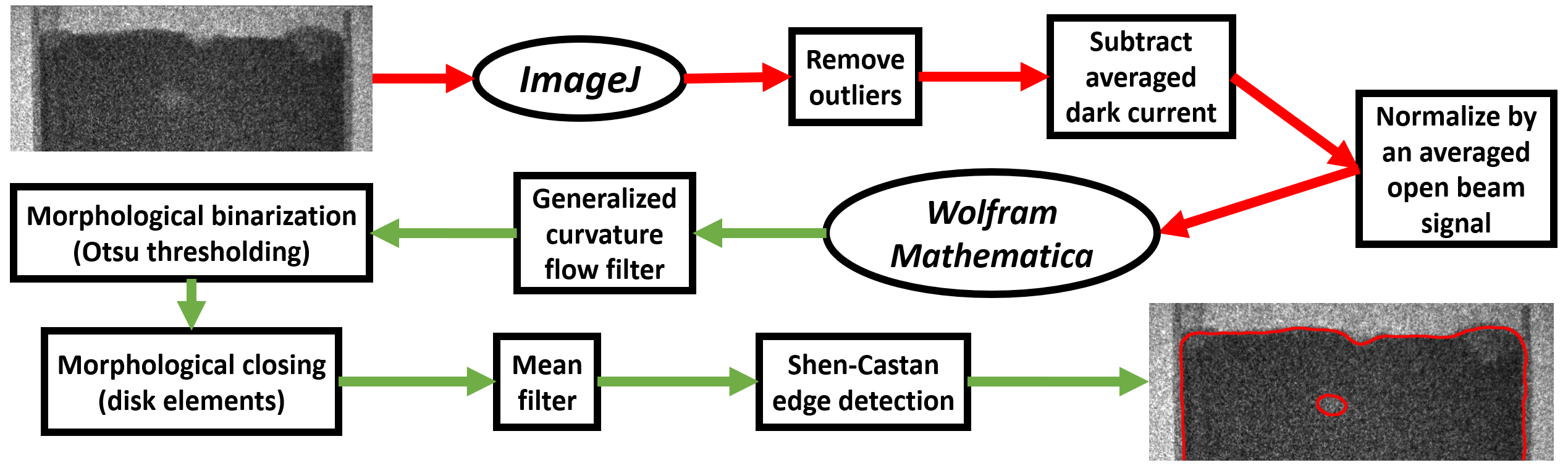}
\caption{The image processing pipeline, from raw images to extracted phase boundary shapes.}
\label{fig:pipeline}
\end{figure}

Several criteria dictated the \emph{Mathematica} code development: reliable bubble and shape detection, minimization of false positive and detection failure rates, bubble detection throughout the entire FOV, robustness and applicability for the lowest possible signal-to-noise ratio (SNR). Additional goals included high enough shape detection precision to enable phase interface (Ar/Ga and Ga/air) tracking and shape strain rate determination. After testing a large number of different options, the authors established that, thus far, the best approach is as it appears in Figure \ref{fig:pipeline}, wherein the constructed image processing pipeline is illustrated.

Curvature flow filter (CFF) was chosen for denoising, since bubble shapes are elliptic and because CFF aggressively erodes sharp edges that are associated with observed image artefacts. CFF diffuses pixel brightness $I$ over a virtual time interval $\tau$ according to Eq. \ref{eq:curvature-flow}, where diffusion rates tangential and normal to edges (localized at $\Delta I = 0$) are regulated by a control function $c(I,k)$:

\begin{equation}
\label{eq:curvature-flow}
\pdv{I}{t} = |\nabla I| \cdot \nabla \left(  c(I,k) \cdot \frac{\nabla I}{|\nabla I|}  \right); ~~t \in [0,\tau]
\end{equation}
where $k$ is a control parameter. Equation \ref{eq:curvature-flow} was solved over image pixels using the finite difference method (FDM).

\begin{figure}[htbp]
\centering
\includegraphics[width=\textwidth]{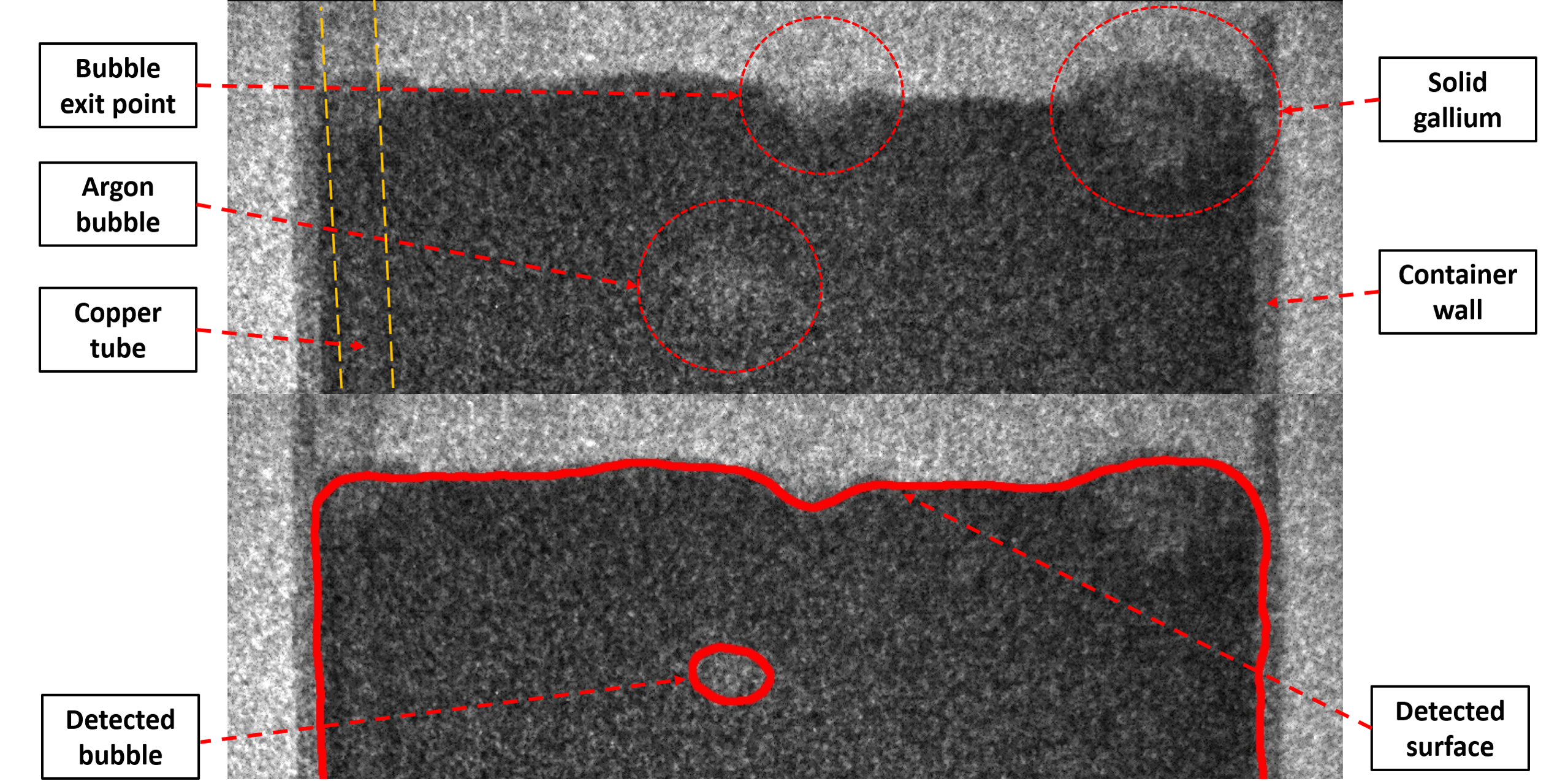}
\caption{(a) A sample pre-processed image with highlighted characteristic features and (b) a post-processed image with derived air/Ga and Ar/Ga interfaces.}
\label{fig:detect}
\end{figure}

After determining the appropriate parameters for all pipeline elements ($c(I,k) = exp(- |\nabla I|^2/k^2)$ was chosen), frames from all measurements were processed. A representative example of input/output is shown in Figure \ref{fig:detect}. Bubble shapes were identified (assumed to be Jordan curves) and analysed, and parameters such as projected area, centroid coordinates and semi axes were determined by fitting ellipses into detected shapes. For this, an additional subroutine consisting of short-range Gaussian blurring, Otsu binarization and morphological thinning was implemented.

Finally, logical and statistical filters were applied to resulting data, removing artefacts left over (if any) from preceding processing stages. Generated output and another set of logical filters were used to compute velocities, trace trajectories and derive parameter correlations.

\section{Simulations}
\label{sec:simulations}

The experiment was modelled numerically to verify that observed effects are not artefacts due to imperfections in the setup. Simulations were carried out in an open-source finite volume method (FVM) package \emph{OpenFOAM} using the volume of fluid (VOF) method, according to incompressible Navier-Stokes equations, a continuity equation and a transport equation for Ga volume fraction, with linear blending functions for density and viscosity. Boundary conditions were (see Figure \ref{fig:setup}b for geometry): zero flow velocity at vessel walls; zero relative pressure at the top opening to allow for gas flow circulation; constant mass flow rate at the tube inlet.

Magnetic field within the Ga container was computed using an open-source finite element method (FEM) package \emph{Elmer} according to magnetic induction equation in terms of magnetic vector potential, and relevant force density contributions (gravity, Lorentz force, surface tension) were passed to \emph{OpenFOAM} via \emph{Elmer-OpenFOAM} (EOF) coupling library \cite{r12}. 

Material physical properties were as follows - Ga density: $6080~kg/m^3$; Ga viscosity: $1.97~mPa \cdot s$; Ga electrical conductivity: $3.70 \cdot 10^6~S/m$; Ga surface tension: $0.72~N/m$; Ar density: $1.784~kg/m^3$; Ar viscosity: $2.30 \cdot 10^{-5}~Pa \cdot s$. 

Preliminary analysis and simulations show that the magnetic Reynolds number is $0.01 < Re_m < 0.1$, so current induced by metal flow through magnetic field should be important. However, in order to accelerate the computations significantly and obtain preliminary results, the authors decided to neglect (for now) the magnetic field produced by induced currents. The hydrodynamic Reynolds number near bubbles is within $10^3 < Re < 10^4$ and $Re \sim 1$ elsewhere, so an appropriate turbulence model must be used. The $k$-equation subgrid-scale (SGS) large eddy simulation (LES) model was chosen to avoid artificial bubble trajectory stabilization due to overestimated turbulent viscosity introduced by a more standard $k$-$\omega$ shear stress transport (SST) model. Estimates indicate that the Eötvös number is $2.1 < Eo < 4.1$, which corresponds to a flow regime wherein bubbles are of slightly oscillating elliptic shapes. 

The dynamic FVM mesh was comprised of roughly $450K$ elements with $2$x refinement at phase interfaces.   Simulations were performed on a computational cluster at the University of Latvia (UL). A simpler shape detection algorithm was applied to CFD simulation results: short range local adaptive binarization, morphological thinning and filling transform, followed by Shen-Castan edge detection and thinning, yielded bubble contours.

\section{Results}
\label{sec:results}

\begin{figure}[htbp]
\centering
\includegraphics[width=\textwidth]{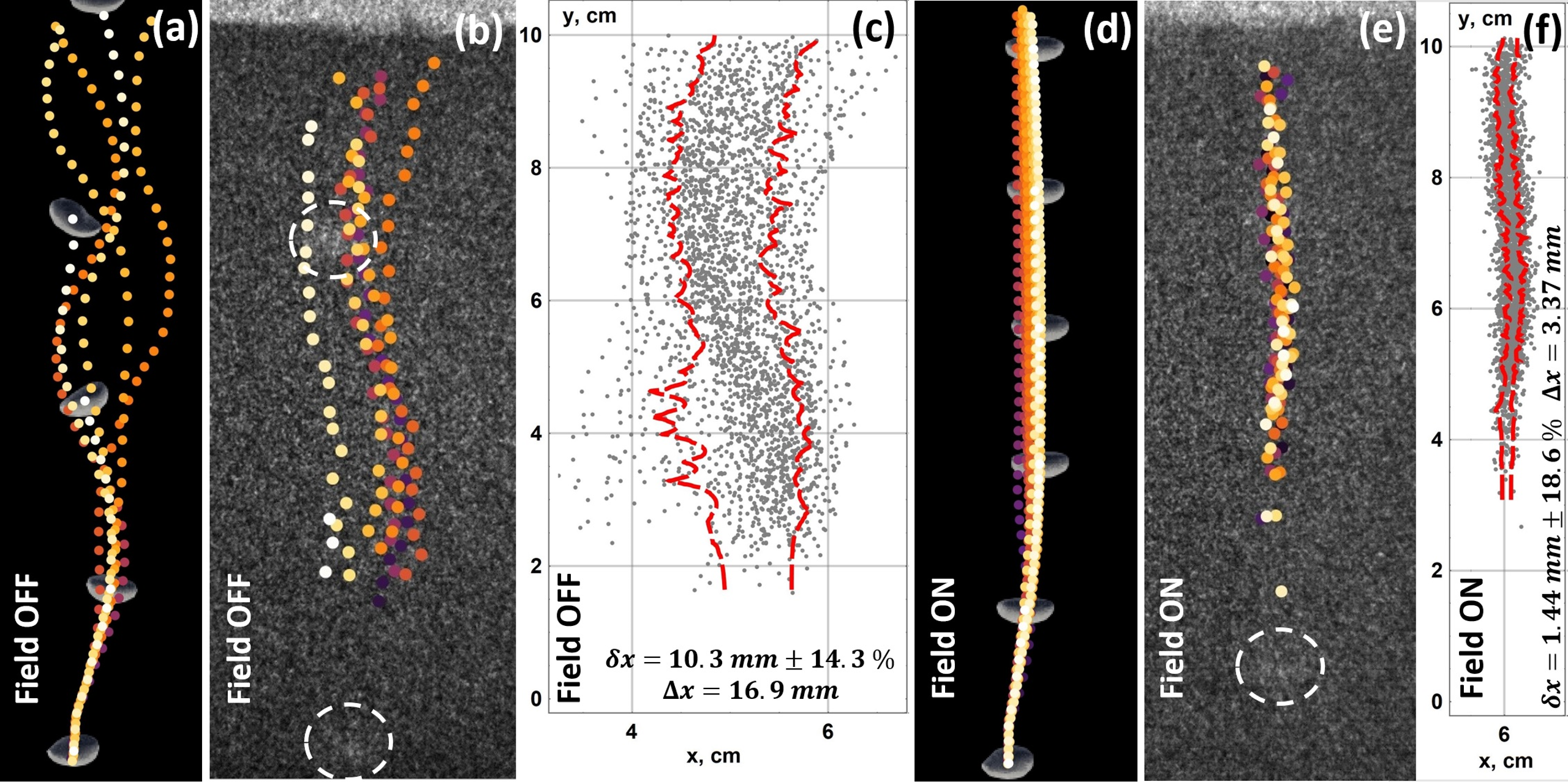}
\caption{Several initial bubble trajectories for a $100 ~cm^3/min$ flow rate derived from simulations (a,d) and experiments (b,e). In cases (a,b) there is no magnetic field, and in (d,e) the field ($ \sim 0.3 ~ T$) is applied. Simulation frame rate matches that of the experiment. Bubble detection points are color coded by order or appearance, dark purple to white. Inlet and free surface are located right beneath and above vertical boundaries of images, respectively. Bubbles are highlighted in experimental images (b,e) by dashed white circles. In (c,f), entire sets of detected bubbles over all frames are shown, without (c) and with (f) horizontal magnetic field. Dashed red lines indicate bubble set envelopes, derived using the statistics-sensitive non-linear iterative peak-clipping (SNIP) algorithm. $\delta x$ and $\Delta x$ in (c,f) stand for mean bubble set envelope thickness and maximum horizontal bubble spread, respectively. Distance scales in (c,f) are identical.}
\label{fig:trajectories}
\end{figure}

\begin{figure}[htbp]
\centering
\includegraphics[width=\textwidth]{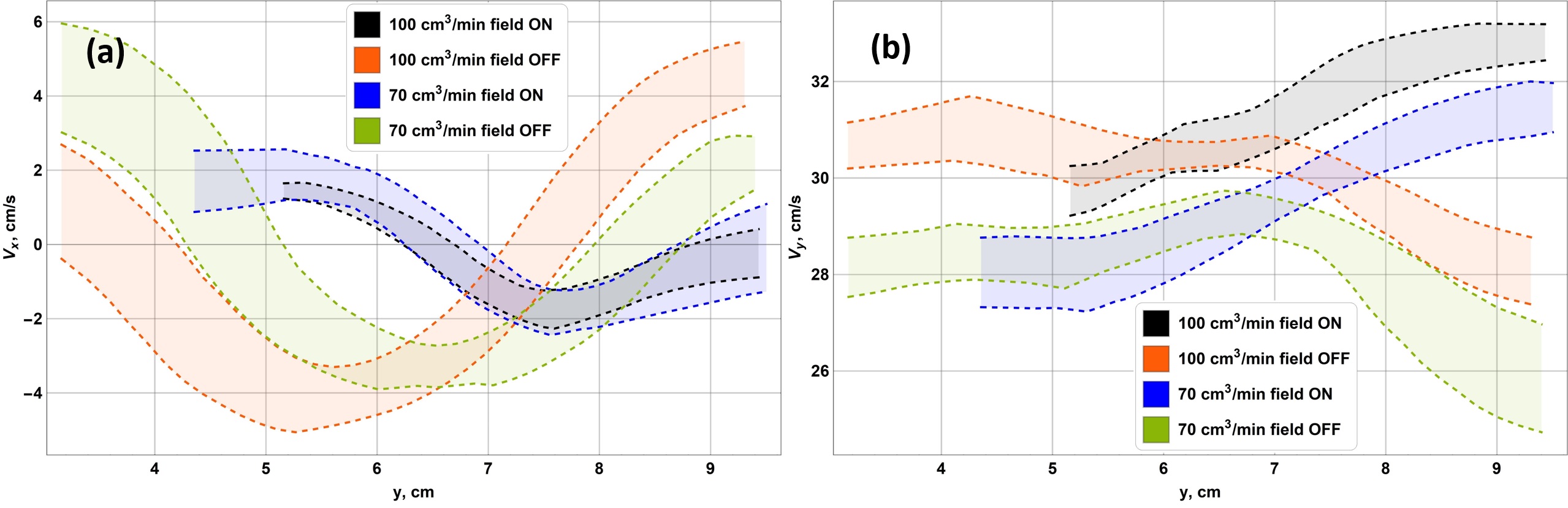}
\caption{Experimentally determined averaged (a) horizontal and (b) vertical velocity components of ascending bubbles at different elevations above the inlet, for different gas flow rates, without and with ($\sim 0.3 ~ T$) applied magnetic field. Colored bands represent averaged curves plus their local errors.}
\label{fig:velocimetry}
\end{figure}

Figure \ref{fig:trajectories} clearly shows the effect of applying magnetic field – bubble trajectory spread is considerably reduced. Stabilization is brought about by electric current due to gallium flow through applied horizontal magnetic field, where flow is induced via fluid displacement by ascending bubbles. Induced current then interacts with magnetic field resulting in the Lorentz force that acts to reduce the velocity component perpendicular to the applied field. This, in turn, leads to bubble wake laminarization, preventing tail vortex detachment stabilizing bubble trajectory \cite{r2}. One can see that simulations and experiments are in very good agreement. The horizontal displacement of individual bubbles is also greatly reduced, as is evident from both Figure \ref{fig:trajectories} and Figure \ref{fig:velocimetry}a. While the horizontal velocity component is reduced (Figure \ref{fig:velocimetry}a), the vertical, conversely, is increased (Figure \ref{fig:velocimetry}b).

These observations are in qualitative agreement with known experimental results obtained by other research groups for similar system dimensionless parameters, which indicates that there are no major issues in the experiment, simulations or the image processing algorithm \cite{r2,r3,r4}. As expected from preliminary analysis, bubble shapes are slightly oscillating, and these shape perturbations are damped when horizontal magnetic field is applied - however, this requires further quantitative analysis. Previously conducted experiments indicate that slight vertical acceleration is expected at $Eo$ values considered herein – this is observed as well.

\section{Conclusions}
\label{sec:conclusions}

The influence of static and rather homogeneous horizontal external magnetic field on the motion of bubbles in an open pool of liquid Ga was investigated both experimentally (using neutron imaging) and \emph{in silico} (using simulations).

High temporal resolution neutron imaging (up to 100 FPS) of dynamics of ascending Ar bubbles (at different gas flow rates) in liquid Ga was performed at the thermal neutron beamline \emph{NEUTRA}. The intrinsically low SNR neutron radiographs were processed via a uniquely tailored image processing pipeline, which enabled successfully perform bubble velocimetry for the resulting images. 

The derived bubble trajectories reveal clear influence of applied magnetic field on bubble velocities, both horizontal and vertical. Significant suppression of horizontal displacement in trajectories is observed for all flow rates, along with reduced trajectory spread, leading to a much more stable flow regime. Vertical acceleration is observed in presence of magnetic field at elevations where the contrary holds without applied field. These observations are in qualitative agreement with known experimental results.

The preliminary CFD simulations also match the experimental results rather well. Statistical data and correlations regarding bubble size, aspect ratio and other parameters have been obtained as well. Thus, we may consider our methodology (the imaging setup and the image processing pipeline) successfully validated.

\section{Acknowledgements}
This work is based on experiments performed at the Swiss spallation neutron source SINQ, Paul Scherrer Institute. The authors would like to thank Jevgenijs Telicko (UL) and Jan Hovind (PSI) for their invaluable assistance with hardware during experiments, and Robert Zboray (Empa Dübendorf, Switzerland), for productive discussions regarding data post-processing. The authors acknowledge the support due to the ERDF project ”Development of numerical modelling approaches to study complex multiphysical interactions in electromagnetic liquid metal technologies” (No. 5 1.1.1.1/18/A/108). This is a preprint for the original paper: \textit{IOS Press, International Journal of Applied Electromagnetics and Mechanics, M. Birjukovs et al., Argon Bubble Flow in Liquid Gallium in External Magnetic Field}, DOI: 10.3233/JAE-209116.


\begin{thebibliography}{99}

\bibitem{r1}
S. Pavlovs, A. Jakovics, E. Baake, V. Sushkovs, Gas bubbles and liquid metal flow influenced by uniform external magnetic field,
\textit{International Journal of Applied Electromagnetics and Mechanics} \textbf{53} (2016), 1-11.

\bibitem{r2}
C. Zhang, Liquid Metal Flows Driven by Gas Bubbles in a Static Magnetic Field,
\textit{PhD thesis} (2009).

\bibitem{r3}
C. Zhang, S. Eckert, G. Gerbeth, Experimental study of single bubble motion in a liquid metal column exposed to a DC magnetic field,
\textit{International Journal of Multiphase Flow} \textbf{31} (2005), 824–842.

\bibitem{r4}
E. Strumpf, Experimental study on rise velocities of single bubbles in liquid metal under the influence of strong horizontal magnetic fields in a flat vessel,
\textit{International Journal of Multiphase Flow} \textbf{97} (2017), 168-185.

\bibitem{r5}
M. Sarma, M. Scepanskis, A. Jakovics, K. Thomsen, R. Nikoluskins, P. Vontobel, T. Beinerts, A. Bojarevics, E. Platacis, Neutron Radiography Visualization of Solid Particles in Stirring Liquid Metal,
\textit{Physics Procedia} \textbf{69} (2015), 457-463.

\bibitem{r6}
E. Baake, T. Fehling, D. Musaeva, T. Steinberg, Neutron radiography for visualization of liquid metal processes: bubbly flow for CO2 free production of Hydrogen and solidification processes in EM field,
\textit{IOP Conference Series: Materials Science and Engineering} \textbf{228} (2017).

\bibitem{r7}
R. Zboray, P. Trtik, 800 fps neutron radiography of air-water two-phase flow,
\textit{Methods X} \textbf{5} (2018), 96-102.

\bibitem{r8}
R. Zboray, P. Trtik, In-depth analysis of high-speed, cold neutron imaging of air-water two- phase flows,
\textit{Flow Measurement and Instrumentation} \textbf{66} (2019), 182-189.

\bibitem{r9}
R. Zboray, V. Dangendorf, I. Mor, B. Bromberger, K. Tittelmeier, Time-resolved Fast Neutron Radiography of Air-water Two-phase Flows,
\textit{Physics Procedia} \textbf{69} (2014).

\bibitem{r10}
R. Zboray, I. Mor, V. Dangendorf, M. Stark, K. Tittelmeier, M. Cortesi, R. Adams, High-frame rate, fast neutron imaging of two-phase flow in a thin rectangular channel,
\textit{Applied Radiation and Isotopes} \textbf{90} (2014), 122–131.

\bibitem{r11}
E. H. Lehmann, P. Vontobel, Properties of the radiography facility NEUTRA at SINQ and its use as European reference facility,
\textit{Nondestructive Testing And Evaluation} \textbf{16(2)} (2001), 191-202.

\bibitem{r12}
J. Vencels, P. Raback, V. Geza, Open-source Elmer FEM and OpenFOAM coupler for electromagnetics and fluid dynamics,
\textit{SoftwareX} \textbf{9} (2019).



\end{thebibliography}
\end{document}